\documentclass[11pt]{article}
\usepackage{graphicx}% Include figure files
\usepackage{epstopdf}
\usepackage{amstext,amsfonts,amsbsy,amssymb,bbm}
\usepackage{pdfsync}
\usepackage{a4wide}
\usepackage{times}
\newcommand{\Ker}{\mathop{\mbox{Ker}}}

\def\Im{\mathop{\rm Im}\nolimits}

\newcommand {\be}[1]{\begin{eqnarray} \mbox{$\label{#1}$}  }
\newcommand{\ee}{\end{eqnarray}}

\newcommand{\pref}[1]{(\ref{#1})}

\newcommand{\iden}{\mathbbm{1}}

\newcommand{\cD}{ {\cal D} }

\newcommand{\cH}{ {\cal H} }

\newcommand{\cP}{ {\cal P} }

\newcommand{\cS}{ {\cal S} }
\newcommand{\cU}{ {\cal U} }

\newcommand{\ga}{ {\alpha} }
\newcommand{\gb}{ {\beta} }

\newcommand{\gd}{ {\delta} }

\newcommand{\gy}{ {\psi} }

\newcommand{\gr}{\rho}

\begin{document}

%%      Final version 
%\newcommand{\be}{\begin{eqnarray}}

%      For draft    %%%%%%%%
%\renewcommand {\be}[1]{
  % {\marginpar{{\scriptsize\ \\ \ #1}}}
       %\begin{eqnarray} \mbox{$\label{#1}$}  }

%\tightenlines

\title{Low rank extremal PPT states and unextendible product bases} 
\author{Jon Magne Leinaas$^a$, Jan Myrheim$^b$ and Per {\O}yvind Sollid$^a$\\ 
${(a)}$ Department of Physics,University of Oslo,\\ P.O.
Box 1048 Blindern, NO-0316 Oslo, Norway\\
${(b)}$ Department of Physics, The Norwegian University of Science and Technology,\\ NO-7491 Trondheim, Norway}

%\date{23 December, 2009}

\maketitle

\begin{abstract}
  It is known how to construct, in a bipartite quantum system, a
  unique low rank entangled mixed state with positive partial
  transpose (a PPT state) from an unextendible product basis (a UPB),
  defined as an unextendible set of orthogonal product vectors.  We
  point out that a state constructed in this way belongs to a
  continuous family of entangled PPT states of the same rank, all
  related by non-singular product transformations, unitary or
  non-unitary.  The characteristic property of a state $\gr$ in such a
  family is that its kernel $\Ker\gr$ has a generalized UPB, a basis
  of product vectors, not necessarily orthogonal, with no product
  vector in $\Im\gr$, the orthogonal complement of $\Ker\gr$.  The
  generalized UPB in $\Ker\gr$ has the special property that it can be
  transformed to orthogonal form by a product transformation.  In the
  case of a system of dimension $3\times 3$, we give a complete
  parametrization of orthogonal UPBs.  This is then a parametrization
  of families of rank 4 entangled (and extremal) PPT states, and we
  present strong numerical evidence that it is a complete
  classification of such states.  We speculate that the lowest rank
  entangled and extremal PPT states also in higher dimensions are
  related to generalized, non-orthogonal UPBs in similar ways.
\end{abstract}

%\pacs{}
%\maketitle

\section{Introduction}

For a composite quantum system, with two separate parts $A$ and $B$,
the mixed quantum states are described by density matrices that can be
classified as being either entangled or separable
(non-entangled). However, there is in general no easy way to classify
a given density matrix as being separable or not. This problem is
referred to as the separability problem, and it has been approached in
the literature in different ways over the past several years
\cite{RPMKHorodecki}. As a part of this discussion there has been a
focus on a subset of the density matrices which includes, but is
generally larger than, the set of separable states. This is the set of
the so-called PPT states, the density matrices that remain positive
under a {\em partial} matrix transposition, with respect to one of the
subsystems, either $A$ or $B$ \cite{Peres96}.

Since it is straight forward to establish whether a density matrix is
PPT, the separability problem is reduced to identifying the subset of
{\em entangled} PPT states. We refer here to the set of separable
states as $\cS$ and the set of PPT states as $\cP$, with
$\cS\subset\cP$. These are both {\em convex} subsets of the full
convex set of density matrices, which we denote as $\cD$, and in
principle the two sets are therefore defined by their extremal
states. The extremal separable states are the pure product states, and
these are also extremal states of the set $\cP$. Since $\cP$ is in
general larger than $\cS$, it has additional extremal states, and
these states are not fully known. The problem of finding and
classifying these additional extremal states is therefore an important
part of the problem to identify the PPT states that are entangled.

We have in two previous publications studied, in different ways, the
problem of finding extremal PPT states in systems of low
dimensions. In \cite{LeinaasMyrheim07} a criterion for extremality was
established and a method was described to numerically search for
extremal PPT states. This method was applied to different composite
systems, and several types of extremal states were found. In a recent
paper \cite{Sollid09} this study has been followed up by a systematic
search for PPT states of different ranks. Series of extremal PPT
states have there been identified and tabulated for different
bipartite systems of low dimensions.

The study in \cite{Sollid09} seems to show that the extremal PPT
states with {\em lowest rank} are somehow special compared to the
other extremal states.  In particular we have found that these density
matrices have no product vectors in their image, but a finite,
complete set of product vectors in their kernel. This was found to be
a common property of the lowest rank extremal PPT states studied
there, for all systems with subsystems of dimensions larger than
2. This property relates these states to a particular construction,
where {\em unextendible product bases}, UPBs for short, are used in a
method to construct entangled PPT states
\cite{Bennett99,DiVincenzo03,Pittenger03a}.

The motivation for the present paper is to follow up this apparent
link between the lowest rank extremal PPT states and the UPB
construction. Our focus is particularly on the rank 4 states of the
3x3 system. The rank 4 extremal PPT states that we find numerically by
the method introduced in \cite{Sollid09} are related by product
transformations to states constructed directly from UPBs. We discuss
this relation and use it to give a parametrization of the rank 4
extremal PPT states.

Although a direct application of the (generalized) UPB construction to
the lowest rank extremal states is restricted to the 3x3 system, the
similarity between these states and the lowest rank extremal states in
higher dimensions indicates that there may exist a generalization of
this construction that is more generally valid. We include at the end
a brief discussion of the higher dimensional cases and only suggest
that a construction method, and thereby a parametrization, of such
states may exist.

%%%%%%%%%%
\section{An extension of the UPB construction of entangled PPT states}

\smallskip
We consider in the following a bipartite quantum system with a Hilbert
space $\cH=\cH_A\otimes\cH_B$ of dimension $N=N_AN_B$. By definition,
a separable state can be written as a density operator of the form
\be{sep}
\gr=\sum_k p_k\,\gy_k\gy_k^{\dag}\;,
\ee
with $p_k\geq 0$, $\sum_k p_k=1$, and with $\gy_k=\phi_k\otimes\chi_k$
as normalized product vectors. The image of $\gr$, $\Im\gr$, is
spanned by these vectors. The fact that $\Im\gr$ must be spanned by
product vectors if $\gr$ is separable is the basis for the UPB
construction, which was introduced in Ref.~\cite{Bennett99}, and used
there to find low-rank entangled PPT states of the 3x3 system. We
review here this construction and discuss a particular generalization.

Consider ${\cal U}$ to be a subspace of 
$\cH$ that is spanned by a set of {\em orthonormal} product vectors
\be{UPB1}
\gy_k=\phi_k\otimes\chi_k\;,\;\;k=1,2,...,p
\ee
which cannot be extended further in $\cH$ to a set of $p+1$ orthogonal
product vectors.  This defines the set as an unextendible product
basis (a UPB). Let ${\cal U}^{\perp}$ be the orthogonal complement to
${\cal U}$. The state proportional to the orthogonal projection onto
${\cal U}^{\perp}$,
\be{Bennett}
\gr_1= a_1\left(\iden - \sum_k \gy_k\gy_k^{\dag}\right),
\ee
with $a_1=1/(N-p)$ as a normalization factor, is then an entangled
PPT state. It is non-separable because $\Im\gr_1={\cal U}^{\perp}$
contains no product vector, and it is PPT because $\gr_1^{\,P}$, the
partial transpose of $\gr_1$ with respect to subsystem $B$, is
proportional to a projection of the same form,
\be{BennettP}
\gr_1^{\,P}= a_1\left(\iden
-\sum_k\tilde{\gy}_k\tilde{\gy}_k^{\,\dag}
\right),
\ee
with $\,\tilde{\gy}_k=\phi_k\otimes\chi_k^{\ast}\,$. The vector
$\chi_k^{\ast}\,$ is the complex conjugate of $\chi_k\,$, in the
same basis in $\cH_B$ as is used for the partial transposition.

The set of product vectors
$\{\,\tilde{\gy}_k=\phi_k\otimes\chi_k^{\ast}\,\}$ is a new
orthonormal UPB, which generally spans a different subspace than the
original set $\{\,\gy_k=\phi_k\otimes\chi_k\,\}$.  However, it may
happen that there exists a basis for the Hilbert space $\cH_B$ of the
second subsystem in which all the vectors $\chi_k$ have real
components.  In such a basis the two UPB sets are identical and the
state $\gr_1$ is PPT for the simple reason that it is invariant under
partial transposition, $\gr_1^{\,P}=\gr_1$.  All the states given as
examples in Ref.~\cite{Bennett99} are of this special kind.

An entangled PPT state $\gr_1$ defined by this UPB construction is a
rather special density operator.  $\Ker\gr_1$ is spanned by product
vectors, while $\Im\gr_1$ contains no product vector.  Since $\gr_1$
is proportional to the orthogonal projection onto the subspace
$\Im\gr_1$, it is the maximally mixed state on this subspace.  There
is also a symmetry between $\gr_1$ and $\gr_1^{\,P}$, such that
$\gr_1^{\,P}$ shares with $\gr_1$ all the properties mentioned above,
and has the same rank $N-p$, where $N=N_AN_B$ is the dimension of the
Hilbert space and $p$ is the number of product vectors in the UPB.

Implicitly the construction implies limits to the rank of $\gr_1$.
Thus, for a given Hilbert space of dimension $N=N_AN_B$ there is a
lower limit to the number of product vectors in a UPB, which follows
from the requirement that there should exist no product vector in the
orthogonal space ${\cal U}^{\perp}$.  The corresponding upper bound on
the rank $m$ of $\gr_1$, as discussed in Ref.~\cite{Sollid09}, is
given by $m<N-N_A-N_B+2$. There is also a lower bound
$m>{\rm max}\{N_A,N_B\}$, which is the general lower bound on the rank
of entangled PPT states with full local rank \cite{Horodecki00}.  In
some special cases there exist more restrictive bounds than the ones
given here~\cite{Alon01}.

For the 3x3 system these two bounds allow only one value $m=4$ for the
rank of a state $\gr_1$ consructed from a UPB, and for this rank
explicit constructions of UPBs exist \cite{Bennett99}.  Also in higher
dimensions a few examples of UPB constructions have been given
\cite{DiVincenzo03}.

The extension of the UPB construction that we shall consider here is
based on a certain concept of equivalence between density operators
previously discussed in \cite{LeinaasMyrheim06}. The equivalence
relation is defined by transformations between density operators of
the form
\be{rho2}
\gr_2=a_2V\gr_1V^{\dag}\;,
\ee
where $a_2$ is a positive normalization factor, and
$V=V_A\otimes V_B$, with $V_A$ and $V_B$ as non-singular linear
operators on $\cH_A$ and $\cH_B$, respectively. The operators $\gr_1$
and $\gr_2$ are equivalent in the sense that they have in common
several properties related to entanglement.  In particular, the form
of the operator $V$ implies that separability as well as the PPT
property are preserved under the transformation
\pref{rho2}. Preservation of separability follows directly from the
product form of the transformation, while preservation of PPT follows
because the partially transposed matrix $\gr_1^{\,P}$ is transformed
in a similar way as $\gr_1$,
\be{rho2P}
\gr_2^{\,P}=a_2\tilde{V}\gr_1^{\,P}\tilde{V}^{\dag}\;,
\ee
with $\tilde{V}=V_A\otimes V_B^{\ast}$.  If $\gr_1$ and $\gr_1^{\,P}$
are both positive then the transformation equations show explicitly
that the same is true for $\gr_2$ and $\gr_2^{\,P}$. Furthermore,
since the operators $V$ and $\tilde{V}$ are non-singular, the ranks of
$\gr_1$ and $\gr_2$ are the same, and also the ranks of $\gr_1^{\,P}$
and $\gr_2^{\,P}$. The same is true for the {\em local} ranks of the
operators, which are the ranks of the reduced density operators,
defined with respect to the subsystems $A$ and $B$. Finally, if
$\gr_1$ is an extremal PPT state, so is $\gr_2$.

Let us again assume $\gr_1$ to be given by the expression
\pref{Bennett}. Since the product operator $V$ is an invertible
mapping from $\Im\gr_1$ to $\Im\gr_2$, and since $\Im\gr_1$ contains
no product vector, there is also no product vector in $\Im\gr_2$,
hence $\gr_2$ is entangled.  Similarly, the product operator
$(V^{\dag})^{-1}$ is an invertible mapping from $\Ker\gr_1$ to
$\Ker\gr_2$, and it maps the UPB in $\Ker\gr_1$, eq.~(\ref{UPB1}),
into a set of product vectors in $\Ker\gr_2$,
\be{UPB2}
\gy'_k
=((V_A^{\dag})^{-1}\phi_k)\otimes
 ((V_B^{\dag})^{-1}\chi_k)\;,\;\;k=1,2,...,p\;.
\ee
If the operators $V_A$ and $V_B$ are both unitary, then this is
another UPB of orthonormal product vectors, and $\gr_2$ is
proportional to a projection, just like $\gr_1$.  More generally,
however, we may allow $V_A$ and $V_B$ to be non-unitary.  Then the
product vectors $\gy'_k$ in $\Ker\gr_2$ will no longer be orthogonal,
but $\gr_2$ is nevertheless an entangled PPT state. It has the same
rank as $\gr_1$, but is not proportional to a projection.

Since the normalization of the density operators $\gr_1$ and $\gr_2$
is taken care of by the normalization factors $a_1$ and $a_2$, we may
impose the normalization condition $\det V_A=\det V_B=1$, which
defines the operators as belonging to the Special Linear (SL) groups
on $\cH_A$ and $\cH_B$.  We will say then that the two density
operators $\gr_1$ and $\gr_2$, related by a transformation of the form
\pref{rho2}, are $\mbox{SL}\otimes\mbox{SL}$ equivalent, or simply SL
equivalent.

The above construction motivates a generalization of the concept of a
UPB, where we no longer require the product vectors to be orthogonal.
This generalization has also previously been proposed in the
literature~\cite{Pittenger03a}.  In the following we will refer to an
unextendible product basis of orthogonal vectors as an
{\em orthogonal} UPB.  A more general UPB is then a set of product
vectors that need not be orthogonal (need not even be linearly
independent), but satisfies still the condition that no product vector
exists in the subspace {\em orthogonal} to the set. The UPB defined by
\pref{UPB2} is a special type of generalized UPB, since it is SL
equivalent to an orthogonal UPB. More general types of UPBs exist, and
they are in fact easy to generate, since an arbitrarily chosen set of
$k$ product vectors is typically a generalized UPB, in the above
sense, when $k$ is sufficiently large. However, if it is {\em not} SL
equivalent to an orthogonal UPB, then we have no guarantee that there
will be any entangled PPT state in the subspace $\cU^{\perp}$
orthogonal to the generalized UPB.

%%%%%%%%%%
\section{Parametrizing the UPBs of the 3x3 system}

We focus now on the orthogonal UPBs in the 3x3 system, which must have
precisely 5 members.  In fact, for any given set of 4 product vectors
$\phi_k\otimes\chi_k$, there exists a product vector $\phi\otimes\chi$
orthogonal to all of them, for example with
$\phi_1\perp\phi\perp\phi_2$ and $\chi_3\perp\chi\perp\chi_4$. And
with 6 members in an orthogonal UPB, it would define a rank 3
entangled PPT state, which is known not to exist \cite{Horodecki00}.

\begin{figure}[h]
\begin{center}
\includegraphics[width=5cm]{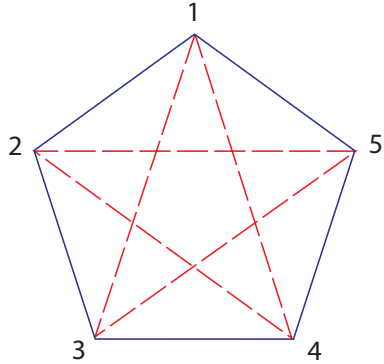}
\end{center}
\caption{\small Diagrammatic representation of the orthogonality
  relations in a 5-dimensional UPB of the 3x3 system. The corners of
  the diagram represent the product vectors of the UPB, and the lines
  represent orthogonality between pairs of states. There are two types
  of orthogonality, represented by the solid (blue) lines and the
  dashed (red) lines. The solid lines represent orthogonality between
  the vectors of the products that belong to subsystem $A$ and the
  dashed lines represent orthogonality between the vectors belonging
  to subsystem $B$. \label{Bennettfig}}
\end{figure}

The general condition for 5 product vectors to form an orthogonal UPB
in the 3x3 system was discussed in Ref.~\cite{Bennett99}.  The
condition implies that for any choice of {\em three} product vectors
from the set, the first factors $\phi_k$ are linearly independent and
so are the second factors $\chi_k$. The orthogonality condition
further implies that if the product vectors are suitably ordered,
there is a cyclic set of orthogonality relations between the factors
of the products, of the form
\be{UPBcond3x3}
\phi_1\perp\phi_2\perp\phi_3\perp
\phi_4\perp\phi_5\perp\phi_1\;,\qquad
\chi_1\perp\chi_3\perp\chi_5\perp
\chi_2\perp\chi_4\perp\chi_1\;.
\ee
In Fig.~\ref{Bennettfig} the situation is illustrated by a diagram
composed of a pentagon and pentagram, where each corner represents a
product vector. Each pair of vectors is interconnected by a line
showing their orthogonality. A solid (blue) line indicates
orthogonality between $\phi$ states (of subsystem $A$) and a dashed
(red) line indicates orthogonality between $\chi$ states. As shown in
the diagram, precisely two A lines and two B lines connect any given
corner with the other corners of the diagram.

%%%%%%%%%%
%%%%%%%%%%

Introducing a complete set of orthonormal basis vectors $\alpha_j$
in $\cH_A$, we write
\be{phiu}
\phi_k=\sum_{j=1}^3 u_{jk}\,\alpha_j\;,\qquad
k=1,2,3,4,5\;.
\ee
We may choose, for example, $\alpha_1$ proportional to $\phi_1$ and
$\alpha_2$ proportional to $\phi_2$.  If we multiply each basis vector
$\alpha_j$ by a phase factor $\omega_j$, and each vector $\phi_k$ by a
normalization factor $N_k$, we change the $3\times 5$ matrix $u_{jk}$
into $\omega_j^{\,-1}N_ku_{jk}$.  It is always possible to choose
these factors so as to obtain a standard form
\be{uvec}
u=\pmatrix{
1 & 0 & a & b & 0\cr
0 & 1 & 0 & 1 & a\cr
0 & 0 & b &-a & 1},
\ee
with $a$ and $b$ as real and strictly positive parameters, and with
the vectors $\phi_k$ not normalized to length $1$.  A similar
parametrization of the vectors of subsystem $B$ with orthonormal
basis vectors $\beta_j$ gives
\be{chiv}
\chi_k=\sum_{j=1}^3 v_{jk}\,\beta_j\;,\qquad
k=1,2,3,4,5\;,
\ee
and a standard form
\be{vvec}
v=\pmatrix{
1 & d & 0 & 0 & c\cr
0 & 1 & 1 & c & 0\cr
0 &-c & 0 & 1 & d},
\ee
with two more positive parameters $c$ and $d$.  Thus, an arbitrary
orthogonal UPB is defined, up to unitary transformations in $\cH_A$
and $\cH_B$, by four continuous, positive parameters $a,b,c,d$.

Note that, for a given UPB, the parameter values are not uniquely
determined, since the above prescription does not specify a unique
ordering of the 5 product vectors within the set.  Any permutation
that preserves the orthogonality relations pictured in
Fig.~\ref{Bennettfig} will generate a new set of values of the
parameters that define the same UPB. These permutations form a
discrete group with 10 elements, generated by the cyclic shift $k\to
k+1$, and the reflection $k\to 6-k$.

Given the orthonormal basis vectors $\alpha_j$ in $\cH_A$ and
$\beta_j$ in $\cH_B$, we may think of the four positive parameters
$a,b,c,d$ as defining not only one single orthogonal UPB, but a
continuous family of generalized UPBs that are SL equivalent to this
particular orthogonal UPB.  The parameter values defining one such
family may be computed from any UPB in the family via SL invariant
quantities, in the following way.  Given the product vectors
$\phi_k\otimes\chi_k$ for $k=1,2,3,4,5$, not necessarily orthogonal,
we introduce expansion coefficients as in \pref{phiu} and arrange them
as column vectors
\be{ukvec}
u_k=\pmatrix{u_{1k}\cr u_{2k}\cr u_{3k}\cr}.
\ee
Then we introduce the following quantities,
\be{inv12}
s_1&\!\!\!=&\!\!\!
-\frac{\det(u_1 u_2 u_4)\,\det(u_1 u_3 u_5)}
      {\det(u_1 u_2 u_5)\,\det(u_1 u_3 u_4)}=a^2\;,
\nonumber\\
s_2&\!\!\!=&\!\!\!
-\frac{\det(u_1 u_2 u_3)\,\det(u_2 u_4 u_5)}
      {\det(u_1 u_2 u_4)\,\det(u_2 u_3 u_5)}=\frac{b^2}{a^2}\;,
\ee
where the values to the right are determined from the parametrization
\pref{uvec} of the orthogonal UPB defining the family.  Similarly, we
define
\be{inv34}
s_3&\!\!\!=&\!\!\!
\frac{\det(v_1 v_2 v_3)\,\det(v_1 v_4 v_5)}
     {\det(v_1 v_2 v_5)\,\det(v_1 v_3 v_4)}=c^2\;,
\nonumber\\
s_4&\!\!\!=&\!\!\!
\frac{\det(v_1 v_3 v_5)\,\det(v_2 v_3 v_4)}
     {\det(v_1 v_2 v_3)\,\det(v_3 v_4 v_5)}=\frac{d^2}{c^2}\;.
\ee
The quantities $s_1,s_2,s_3,s_4$ defined in terms of $3\times 3$
determinants are useful because they are invariant under SL
transformations as in~\pref{UPB2}, and in addition they are
independent of the normalization of the column vectors $u_k$ and
$v_k$.  Obviously, many more similar invariants may be defined from 5
product vectors, but these four invariants are sufficient to
characterize a family of UPBs that are SL equivalent to an orthogonal
UPB.

There exists a less obvious further extension of the set of
invariants.  In fact, there are always 6 vectors that can be used to
define invariants, since in addition to the 5 linearly independent
product vectors of the UPB, the space spanned by these will always
contain a 6th product vector.  In the case of an orthogonal UPB, given
by the parameters $a,b,c,d$, we have found (by means of a computer
algebra program) explicit polynomial expressions for the components of
the one extra product vector.  We have checked, both analytically and
numerically, that the existence of exactly 6 product vectors is a
generic property of a 5 dimensional subspace of the $3\times 3$
dimensional Hilbert space $\cH$.  This number of product vectors is
also consistent with the formula discussed in \cite{Sollid09}, which
specifies more generally, as a function of the dimensions, the number
of product vectors in a subspace of $\cH$. For an orthogonal UPB in
the 3x3 system the 6th vector is singled out because it is not
orthogonal to the other vectors, but for a non-orthogonal UPB there is
no intrinsic difference between the 6 vectors of the set, which should
therefore be treated on an equal footing.

For a UPB that is SL equivalent to an orthogonal UPB there are strong
restrictions on the values of invariants of the above kind, since they
are all rational functions of the four real parameters $a,b,c,d$. In
particular, they must all take real values.  A given choice of four
invariants, as in \pref{inv12} and \pref{inv34}, is sufficient to
define the parameter space for the equivalence classes of these UPBs.
But since the 6 product vectors listed in any order define the same
UPB, and the same PPT state, there is a discrete set of $6!=720$
symmetry transformations that introduce identifications between points
in the corresponding parameter space.  As we shall see below, the
requirement that all four invariants $s_1,s_2,s_3,s_4$ should be
positive allows 60 different orderings from the total of 720.

One should note that for a generalized UPB consisting of 5 randomly
chosen product vectors the invariants will in general be complex
rather than real, and it is not a priori clear that four invariants
are sufficient to parametrize the equivalence classes of random UPBs.

%%%%%%%%%%
\section{Classifying the rank 4 entangled PPT states}

We have in \cite{Sollid09} described a method to generate PPT states
$\gr$ for given ranks $(m,n)$ in low-dimensional systems, with
$m={\rm rank}\,\gr$ and $n={\rm rank}\,\gr^P$. By repeatedly using
this method with different initial data we have generated a large
number of different PPT states of rank $(4,4)$ in the 3x3 system.
They are all entangled PPT states, and as a consequence they are
extremal PPT states. This follows from the fact that if they were not
extremal they would have to be convex combinations involving entangled
PPT states of even lower ranks, and such states do not exist.

The remarkable fact is that every one of these states has a UPB in its
kernel which is SL equivalent to an {\em orthogonal} UPB, and the
state itself is SL equivalent to the state constructed from the
orthogonal UPB.  We regard our numerical results as strong evidence
for our belief that the four real parameters which parametrize the
orthogonal UPBs give a complete parametrization of the rank 4
entangled PPT states of the $3\times 3$ system, up to the SL (or more
precisely $\mbox{SL}\otimes\mbox{SL}$) equivalence. We will describe
here in more detail the numerical methods and results that lead us to
this conclusion.
 
Assume $\gr$ to be an entangled PPT state of rank $(4,4)$, found by
the method described in \cite{Sollid09}.  The question to examine is
whether it is SL equivalent to an entangled PPT state defined by the
orthogonal UPB construction. We therefore make the {\em ansatz} that
it can be written as $\gr\equiv\gr_2=a_2V\gr_1V^{\dag}$, where $\gr_1$
is defined by a so far unknown {\em orthogonal} UPB, parametrized as
in \pref{uvec} and \pref{vvec}, and where the transformation $V$ is of
product form, $V=V_A\otimes V_B$.  We consider how to compute the
product transformation $V$, assuming that it exists.  The fact that we
are able to find such a transformation for every $(4,4)$ state is a
highly non-trivial result.

Given $\gr$, the first step is to find all the product vectors in
$\Ker\gr$.  We solve this as a minimization problem: a normalized
product vector $\psi=\phi\otimes\chi$ with $\gr\psi=0$ is a minimum
point of the expectation value $\psi^{\dag}\gr\psi$.  Details of the
method we use are given in Ref.~\cite{Sollid09}.  Empirically, we
always find exactly 6 such product vectors
$\psi_k=\phi_k\otimes\chi_k$, $k=1,2,\ldots,6$, any 5 of which are
linearly independent and form a UPB, typically non-orthogonal.

Although the numbering of the 6 product vectors is arbitrary at this
stage, we compute the invariants $s_1,s_2,s_3,s_4$, substituting
$\phi_k$ for $u_k$ and $\chi_k$ for $v_k$, with $k=1,2,\ldots,5$.  As
shown by the previous discussion all the four invariants have to be
real, otherwise no solution can exist.  A random UPB has complex
invariants, and the empirical fact the invariants are always real for
a UPB in $\Ker\gr$ where $\gr$ is a rank $(4,4)$ entangled PPT state,
is a non-trivial test of the hypothesis that such a UPB can be
transformed into orthogonal form.

It is not sufficient that the invariants are real. As shown by the
expressions \pref{inv12} and \pref{inv34} there has to exist an
ordering of the product vectors where all the four invariants take
positive values.  The signs of the invariants will depend on the
ordering of the product vectors, and most orderings produce both
positive and negative invariants.  For the rank $(4,4)$ density
matrices that we have constructed, it turns out that it is always
possible to renumber the 5 first vectors in the set in such a way that
all four invariants become positive.  This is a further non-trivial
test of our hypothesis.

There are in fact, in all the cases we have studied, precisely $10$ of
the $5!$ permutations of the 5 vectors that give positive values of
the four invariants. This means that such an ordering is unique up to
the symmetries noticed for the diagram in
Fig.~\ref{Bennettfig}. However, there is a further symmetry, since the
reordering which gives positive invariants works for any choice of the
6th vector of the set. The possible reorderings of all 6 product
vectors which preserve the positivity of the invariants therefore
define a discrete symmetry group with altogether $6\times 10=60$
elements, which defines mappings between different, but equivalent,
representations of the UPB in terms of the set of four real and
positive invariants.  The corresponding parameter transformations are
given in the Appendix.

Assume now, for a given rank $(4,4)$ state, that a ``good'' numbering
has been chosen for the 6 product vectors $\psi_k=\phi_k\otimes\chi_k$
in the corresponding UPB, so that the four invariants defined by the
first 5 vectors are all real and positive.  The problem to be solved
is then to find the transformation that brings the UPB into orthogonal
form. This means to find $3\times 3$ matrices $C$ and $D$ such that
$\phi_k=N'_kCu_k$ and $\chi_k=N''_kDv_k$ for $k=1,2,\ldots,5$, with
unspecified normalization constants $N'_k$ and $N''_k$.  Here the
vectors $u_k$ and $v_k$ belong to an orthogonal UPB as given by the
equations~(\ref{uvec}) and~(\ref{vvec}), and these vectors are all
known at this stage, because the invariants $s_1,s_2,s_3,s_4$
determine the parameters $a,b,c,d$. The transformation matrices $C$
and $D$ correspond to $V_A^\dag$ and $V_B^\dag$ in \pref{UPB2}.  The
condition for two vectors $\phi_k$ and $Cu_k$ to be proportional is
that their antisymmetric tensor product vanishes, hence we write the
following homogeneous linear equations for the matrix $C$,
\be{eqC}
\phi_k\wedge(Cu_k)=\phi_k\otimes(Cu_k)-(Cu_k)\otimes\phi_k=0\;,
\qquad k=1,2,\ldots,5\;.
\ee
Since the antisymmetric tensor product $\phi_k\wedge(Cu_k)$ has, for
given $k$, 3 independent components, there are altogether 15 linear
equations for the 9 unknown matrix elements $C_{ij}$.  We may
rearrange the $3\times 3$ matrix $C$ as a $9\times 1$ matrix $C$ and
write a matrix equation
\be{MCiszero}
MC=0\;,
\ee
where $M$ is a $15\times 9$ matrix.  This equation implies that
$(M^{\dag}M)C=0$.  The other way around, the equation $(M^{\dag}M)C=0$
implies that $(MC)^{\dag}(MC)=C^{\dag}(M^{\dag}M)C=0$ and hence
$MC=0$.  Thus we may compute the matrix $C$ as an eigenvector with
zero eigenvalue of the Hermitean $9\times 9$ matrix $M^{\dag}M$.  The
matrix $D$ is computed in a similar way.

It is a final non-trivial empirical fact for the $(4,4)$ states we
have found, that solutions always exist for the matrices $C$ and $D$,
whenever the ordering of the 6 product vectors
$\psi_k=\phi_k\otimes\chi_k$ is such that the invariants
$s_1,s_2,s_3,s_4$ are positive.

The result is that every rank $(4,4)$ state of the 3x3 system which we
have found in numerical searches \cite{Sollid09} can be transformed
into a projection operator with an orthogonal UPB in its kernel. We
have also checked the published examples of entangled PPT states of
rank $(4,4)$, which are based on special constructions
\cite{Bennett99,DiVincenzo03,BrussPeres00,Ha03}, and have got the same
result for these states.  The explicit transformations have been found
numerically by the method discussed above, and in all cases the four
parameters $a,b,c,d$ have been determined, with values that are unique
up to arbitrary permutations of product vectors from the $60$ element
symmetry group.

%\pagebreak

%%%%%%%%%%
\section{Summary and outlook}

The main result of this paper is a classification of the rank 4
entangled PPT states of the 3x3 system.  We find empirically that
every state of this kind is equivalent, by a product transformation of
the form $\mbox{SL}\otimes\mbox{SL}$, to a state constructed from an
orthogonal unextendible product basis.  We refer to this type of
equivalence as SL equivalence.  We have shown how to parametrize the
orthogonal UPBs by four real and positive parameters, and we have
described how permutations of the vectors in the UPB give rise to
identifications in the four-parameter space.

The concept of SL equivalence of states and of product vectors leads
to a generalization of the concept of unextendible product bases so as
to include sets of non-orthogonal product vectors, and further to the
concept of equivalence classes of generalized UPBs that are SL
equivalent to orthogonal UPBs.  Thus, the parametrization of the
orthogonal UPBs by four positive parameters is at the same time a
parametrization of the corresponding equivalence classes of
generalized UPBs.

We have described a method for checking whether a given rank 4
entangled PPT state in the 3x3 system is equivalent, by a product
transformation, to a state constructed from an orthogonal UPB.  It is
a highly non-trivial result that all the rank-four entangled states
that we have produced numerically, and all states of this kind that we
have found in the literature, are SL equivalent to states that are
generated from orthogonal UPBs.  This we take as a strong indication
that the parametrization of the UPBs in fact gives also a
parametrization of all the equivalence classes of rank 4 entangled PPT
states of the 3x3 system.

Apart from the pure product states, the rank 4 entangled PPT states
are the lowest rank {\em extremal} PPT states among the 3x3 states
that we have found in numerical searches, as reported on in
\cite{Sollid09}. The property of such a state, that it has a
non-orthogonal UPB in its kernel, which means that there is a complete
set of product vectors in $\Ker\gr$ and no product vector in $\Im\gr$,
is shared with the lowest rank extremal PPT states of the other
systems that we have studied, of dimensions different from 3x3. This
has led us to conjecture that this is a general feature of the lowest
rank extremal PPT states, valid also in higher dimensional systems
\cite{Sollid09}, and to speculate that there may exist a
generalization of the construction used for the 3x3 system in terms of
orthogonal UPBs and SL transformations, which can be applied in the
higher dimensional systems.

In higher dimensions the orthogonality condition is harder to satisfy,
and therefore another condition may take its place as the defining
characteristic of a special subset of extremal states from each SL
equivalence class. This hypothetical new condition may involve the
full set of product vectors in the subspace, rather than an
arbitrarily selected subset as in the definition of the orthogonal
UPBs. To examine this possibility, with the aim of parametrizing the
lowest rank extremal PPT states more generally, we consider an
interesting task for further research, and we are currently looking
into the problem.

%\pagebreak
%\vspace*{2\baselineskip}

\subsection*{\bf\large Acknowledgment}

%\noi
%{\bf\large Acknowledgment}\\
Financial support from the Norwegian Research Council is gratefully
acknowledged.

\pagebreak

\appendix

%%%%%%%%%%
\section{Equivalent orderings of the 6 product vectors}

Assume that the sequence of product vectors
$\psi_k=\phi_k\otimes\chi_k$, $k=1,2,3,4,5$, in this order, is
charcterized by parameter values $a,b,c,d$, as computed from the
invariants $s_1,s_2,s_3,s_4$.  It is convenient here to replace the
parameters $a,b,c,d$ by $\ga=a^2$, $\gb=b^2$, $\gamma=c^2$, $\gd=d^2$.

Then the cyclic permutation $\psi_k\mapsto\tilde{\psi}_k$ with
$\tilde{\psi}_1=\psi_5$ and $\tilde{\psi}_k=\psi_{k-1}$ for
$k=2,3,4,5$ corresponds to the following parameter transformation,
which is periodic with period 5,
\be{paramtr1}
\tilde{\ga}&\!\!\!=&\!\!\!\frac{\gb}{1+\ga}\;,
\nonumber\\
\tilde{\gb}&\!\!\!=&\!\!\!\frac{\gb}{\ga(1+\ga)}\;,
\nonumber\\
\tilde{\gamma}&\!\!\!=&\!\!\!\frac{1}{\gamma+\gd}\;,
\\
\tilde{\gd}&\!\!\!=&\!\!\!\frac{\gamma(1+\gamma+\gd)}{\gd(\gamma+\gd)}\;.
\nonumber
\ee
%

%The inversion $\psi_k\mapsto\tilde{\psi}_k=\psi_{6-k}$ for
%$k=1,2,3,4,5$ corresponds to the parameter transformation
%%
%\be{paramtr2}
%\tilde{\ga}&\!\!\!=&\!\!\!\frac{\gb}{1+\ga}\;,
%\nonumber\\
%\tilde{\gb}&\!\!\!=&\!\!\!\frac{\ga(1+\ga+\gb)}{1+\ga}\;,
%\nonumber\\
%\tilde{\gamma}&\!\!\!=&\!\!\!\frac{1}{\gamma+\gd}\;,
%\\
%\tilde{\gd}&\!\!\!=&\!\!\!\frac{\gd}{\gamma(\gamma+\gd)}\;.
%\nonumber\\
%\ee

The inversion $\psi_1\mapsto\tilde{\psi}_1=\psi_1$,
$\psi_k\mapsto\tilde{\psi}_k=\psi_{7-k}$ for
$k=2,3,4,5$ corresponds to the parameter transformation
$\tilde{\ga}=\ga$, $\tilde{\gamma}=\gamma$,
\be{paramtr2}
\tilde{\gb}&\!\!\!=&\!\!\!\frac{\ga(1+\ga)}{\gb}\;,
\nonumber\\
\tilde{\gd}&\!\!\!=&\!\!\!\frac{\gamma(1+\gamma)}{\gd}\;.
\ee

Let $\psi_6=\phi_6\otimes\chi_6$ be the 6th product vector in the 5
dimensional subspace spanned by the above 5 product vectors.  Then the
sequence $\tilde{\psi}_1=\psi_6$, $\tilde{\psi}_2=\psi_5$,
$\tilde{\psi}_3=\psi_3$, $\tilde{\psi}_4=\psi_4$,
$\tilde{\psi}_5=\psi_2$ corresponds to the parameter transformation
$\tilde{\ga}=\gamma$, $\tilde{\gamma}=\ga$,
\be{paramtr3}
\tilde{\gb}&\!\!\!=&\!\!\!\frac{\gb(1+\gamma)((\ga+\gb)(\gamma+\gd)+\gd)}
{\ga(1+\ga+\gb)\gd+(1+\ga)(\ga+\gb)(1+\gamma)}\;,
\nonumber\\
\tilde{\gd}&\!\!\!=&\!\!\!
\frac{(1+\ga)(\gb\gd+(\ga+\gb)\gamma(1+\gamma+\gd))}
{(1+\ga+(1+\ga+\gb)(\gamma+\gd))\gd}\;.
\ee
It is not easy to see by looking at the formulae that this parameter
transformation is its own inverse.

Altogether, these transformations generate a transformation group of
order 60 (with 60 elements), isomorphic to the symmetry group of a
regular icosahedron with opposite corners identified.  Equivalently,
it is the group of proper rotations of the icosahedron, excluding
reflections.  The icosahedron has 12 corners, and we may associate the
6 product vectors with the 6 pairs of opposite corners.

\pagebreak

%%%%%%%%%

%%%%%%%%%%%%

%%%%%%%%%%
\end{document}